\documentclass[aps,twocolumn]{revtex4-1}
\usepackage{graphicx}
\usepackage{epsfig}
\usepackage{amsmath}
\usepackage{latexsym}
\usepackage{amsmath}
\usepackage{amssymb}
\usepackage{float}
\usepackage{epstopdf}
\usepackage{hyperref}
\usepackage{multirow}
\usepackage{xcolor}
\hypersetup{urlcolor=red,citecolor=blue}
\usepackage{bm}
\usepackage{url}

\begin{document}
\title{Heavy baryon spectrum on lattice with NRQCD bottom and HISQ
lighter quarks}
\author{Protick Mohanta}
\author{Subhasish Basak}
\affiliation{School of Physical Sciences, National Institute of
Science Education and Research, HBNI, Odisha 752050, India}

\date{\today}


\begin{abstract}
We determine the mass spectra of heavy baryons containing one or more
bottom quarks along with their hyperfine splittings and various mass
differences on MILC 2+1 Asqtad lattices at three different lattice
spacings. NRQCD action is used for bottom quarks whereas relativistic
HISQ action for the lighter up/down, strange and charm quarks. We
consider all possible combinations of bottom and lighter quarks to
construct the bottom baryon operators for the states $J^P=1/2^+$ and
$3/2^+$.

\end{abstract}

\maketitle

\section{Introduction}

Lattice QCD has been extensively employed to study $B$ physics
phenomenology, especially the decay constants and mixing parameters
needed for CKM matrix elements and the mass differences in the meson
sector \cite{cdavies}. The $B$ mesons spectroscopy and mass splittings
have undergone thorough investigations on lattice, see \cite{cdavies}
and references therein, with increasing impact on heavy flavor
phenomenology, see for instance \cite{gamiz}. However, studying heavy
baryons with bottom quark(s) on lattice is relatively a recent pursuit.
Some of the early studies of heavy baryons on lattice can be found in
\cite{alikhan, woloshyn, lewis, mathur, heechang}. Of late, a slew of
low lying $J^P=1/2^+$ bottom baryons, such as $\Lambda_b$, $\Sigma_b$,
$\Xi_b^\prime$ and $\Omega_b$, have made entries in PDG \cite{pdg}.
Possibilities of discoveries of $J^P=3/2^+$ are rather high, whereas
doubly and triply bottom baryons are right now beyond the reaches of
present experiments. In this state, lattice QCD can provide an insight
into the masses, mass splittings and other properties of such bottom
baryons from the first principle. To this end, quite a few lattice
investigations of heavy baryons containing one, two or three bottom
quarks have been undertaken using a range of light quark actions
\cite{meinel, burch, brown}. For an extensive list on contemporary
lattice literature on heavy baryon see \cite{brown}.

These studies on heavy hadrons with bottom quark(s) are largely made
possible by the use of nonrelativistic QCD action, proposed and
formulated in \cite{lepage, thacker}, because of the well-known fact
that the current lattice spacings, even for as low as 0.045 fm, render
$am_b \gtrsim 1$. Although almost all of the above studies employed
different heavy quark actions for charm quark, HISQ action \cite{hisq}
is becoming an increasingly popular choice for the charm quark. This
approach of simulating bottom quark with NRQCD and the rest of the
quarks {\em i.e.} charm, strange and up/down with HISQ for calculation
of bottom baryon spectra has been adopted in this work.

In this paper, we present our lattice QCD results of heavy baryons involving
one and two bottom quarks. We consider all possible combinations of bottom
quark(s) with charm, strange and up/down {\em lighter} quarks of the form
$(lbb)$, $(llb)$ and $(l_1l_2b)$, where $b$ is the bottom quark and $l$
are the lighter charm, strange and up/down quarks. We are addressing  the
charm quark as ``light" quark in the sense that we have used relativistic
action for it. The action for the lighter quarks is HISQ \cite{hisq} and
NRQCD \cite{lepage, thacker} for the bottom. We discuss these actions in
Section \ref{sec_action}. The propagators generated using nonrelativistic
and relativistic actions are required to be combined to construct baryon
states of appropriate quantum numbers. A discussion to achieve this
combination is spread over both Section \ref{sec_action} and
\ref{sec_operator}. The bottom baryon operators are described in
details in Section \ref{sec_operator}. In the following Section
\ref{sec_simul}, we present the simulation details including the lattice
ensembles used, various parameters and tuning of different quark masses.
The lattice calculations are carried out at three different lattice
spacings with fixed $m_{u/d}/m_s$ value and several quark masses. We
assemble our bottom baryon spectrum results along with hyperfine and
various other mass splittings in the Section \ref{sec_result}. Finally
we conclude and summarize in Section \ref{sec_sum}, which also includes
a comparison of our results to the existing ones.

\section{Quark Actions} \label{sec_action}
As of now the bottom ($b$) quark masses are not small {\em i.e.}
$am_b \nless 1$ in units of the lattice spacings available. The use of
improved NRQCD is the action of choice for the $b$-quarks. We have used  
$\mathcal{O}(v^6)$ NRQCD action in this paper. The charm ($c$) quark
is also similarly heavy enough for existing lattices, but Fermilab
proposal \cite{fermilab} made it possible to work with relativistic
actions for $c$-quark, provided we trade the pole mass with the
kinetic mass. Subsequently, HISQ action \cite{hisq} became available
for relativistic $c$-quark. In this paper, we choose HISQ action
for the $c$-quark along with $s$ and $u/d$ quarks. In this work, as
because we use the same relativistic HISQ action for all quarks except
the bottom, we use the word {\em light} quarks to refer to $c$, $s$ and
$u/d$ quarks. This is similar to what has been done in \cite{eric} for
$B$ meson states calculation. Besides, one of the big advantages of
this choice of action is the ability to use the MILC code \cite{milc_code}
for this bit.

\subsection{NRQCD Action and $b$ quark}
In order to perform lattice QCD computation of hadrons containing bottom
quarks in publicly available relatively coarse lattices, NRQCD
\cite{lepage, thacker} is perhaps the most suitable and widely used quark
action for bottom. As is understood, the typical velocity of a $b$ quark
inside a hadron is nonrelativistic. Comparison of masses of bottomonium
states to the mass of $b$ quark supports the fact that the velocity of
$b$ quark inside hadron ($v^2 \sim 0.1$) is much smaller than the bottom
mass. For example $M_\Upsilon=9460$ MeV whereas $2\times m_b=
8360$ MeV in the $\overline{MS}$ scheme. For bottom hadrons containing
lighter valence quarks, the velocity of the bottom quark is even smaller.
This allows us to study the $b$ quark with nonrelativistic effective
field theory. NRQCD will remain action of choice for $b$ quark until
finer lattices with $am_b < 1$ become widely available.

In NRQCD, the upper and lower components of the Dirac spinor decouple
and the $b$ quark is described by two component spinor field, denoted
by $\psi_h$. NRQCD Lagrangian has the following form
\begin{equation}
\mathcal{L}=\psi_h^{\dagger}(\vec{x},t)[U_4(x)\psi_h(\vec{x},t+1)-
\psi_h(\vec{x},t)+aH\psi_h(\vec{x},t)] 
\end{equation}
where $a$ is the lattice spacing and $U_4(x)$ is the temporal gauge
link operator. $H=H_0+\delta H$ is the NRQCD Hamiltonian where,
\begin{equation}
H_0=-{\tilde{\Delta}^2 \over 2 m_b}-{a\over 4n}{(\Delta^2)^2\over 4 m_b^2}
\hspace{0.15in}\text{and} \;\;\;\; \delta H=\sum_i\delta H^{(i)}
\label{nrqcd_h0}
\end{equation}
The $H_0$ is the leading $\mathcal{O}(v^2)$ term, the $\mathcal{O}(v^4)$
and $\mathcal{O}(v^6)$ terms are in $\delta H$ with coefficients $c_1$
through $c_7$.
\begin{eqnarray}
\delta H^{(1)} &=& -c_1\, {(\Delta^2)^2\over 8 m_b^3} \nonumber\\
\delta H^{(2)} &=&  c_2\, {ig\over 8 m_b^2}\, \left(\vec{\Delta}^{\pm}
\cdot \vec{E} - \vec{E} \cdot \vec{\Delta}^{\pm} \right) \nonumber\\
\delta H^{(3)} &=& -c_3\, {g\over 8 m_b^2}\; \vec{\sigma} \cdot \left(
\tilde{\vec{\Delta}}^{\pm}\times\tilde{\vec{E}}-\tilde{\vec{E}}\times
\tilde{\vec{\Delta}}^{\pm} \right) \nonumber\\
\delta H^{(4)} &=& -c_4\, {g\over 2m_b}\; \vec{\sigma} \cdot \tilde{\vec{B}}
\nonumber\\
\delta H^{(5)} &=& -c_5\, {g\over 8m_b^3}\, \left\{\Delta^2,\vec{\sigma}
\cdot \vec{B}\right\} \nonumber\\
\delta H^{(6)} &=& -c_6\, {3g\over 64m_b^4}\, \left\{\Delta^2,\vec{\sigma}
\cdot \left(\vec{\Delta}^{\pm}\times\vec{E}-\vec{E}\times
\vec{\Delta}^{\pm} \right) \right\} \nonumber \\
\delta H^{(7)} &=& -c_7\, {ig^2\over 8m_b^3}\; \vec{\sigma} \cdot \vec{E}
\times\vec{E} \label{ham}
\end{eqnarray}
The $b$ quark propagator is generated by the time evolution of this
Hamiltonian,
\begin{eqnarray}
\lefteqn{\hspace{-0.5in} G(\vec{x},t+1;0,0) = }\nonumber \\
 & &  \left(1-{aH_0\over 2n}\right)^n \, \left(1-{a\delta H\over 2}\right)
\, U_4(\vec{x},t)^{\dagger} \times \nonumber\\
 & & \left( 1 - {a\delta H\over 2}\right) \, \left(1-{aH_0\over 2n}\right)^n
\, G(\vec{x},t;0,0) \label{green}
\end{eqnarray}
with
\[
G(\vec{x},t;0,0)= \left\{ \begin{array}{cl} 0 & \hspace{0.15in} \text{for}
\;\; t<0 \\ \delta_{\vec{x},0} & \hspace{0.15in} \text{for} \;\; 
t=0 \end{array}\right.
\] 
The tree level value of all the coefficients $c_1$, $c_2$, $c_3$, $c_4$,
$c_5$, $c_6$ and $c_7$ is 1. Here $n$ is the factor introduced to ensure
numerical stability at small $am_b$ \cite{thacker}, where $n > 3/2m_b$.
The symmetric derivative $\Delta^{\pm}$ and Laplacian $\Delta^2$ in
terms of forward and backward covariant derivatives are
\begin{eqnarray}
a \Delta^{+}_{\mu}\psi(x) &=& U_{\mu}(x)\psi(x+a\hat{\mu})-\psi(x)
\nonumber\\
a \Delta^{-}_{\mu}\psi(x) &=& \psi(x)-U^{\dagger}_{\mu}(x-a\hat{\mu})
\psi(x-a\hat{\mu}) \nonumber \\
\Delta^{\pm} &=& {1\over2}(\Delta^+ + \Delta^-) \nonumber \\
\Delta^2 &=& \sum_i\Delta^+_i \Delta^-_i = \sum_i\Delta^-_i
\Delta^+_i  \label{nrqcd_deriv}
\end{eqnarray}
By Taylor expanding the symmetric derivative and the Laplacian operator,
we can find their forms corrected up to $\mathcal{O}(a^4)$ \cite{lepage}
that are used in the above Equation (\ref{ham}).
\begin{eqnarray}
\tilde{\Delta}^{\pm}_i &=& \Delta^{\pm}_i - {a^2\over6}\,\Delta^{+}_i
\Delta^{\pm}_i\Delta^{-}_i \nonumber \\
\tilde{\Delta}^2 &=& \Delta^2 - {a^2\over12} \sum_i \left[\Delta^{+}_i
\Delta^{-}_i\right]^2. \label{impderiv}
\end{eqnarray}
In the same way, the gauge fields are improved to $\mathcal{O}(a^4)$ using
cloverleaf plaquette,
\begin{eqnarray}
a \Delta^{+}_{\rho}F_{\mu\nu}(x) &=& U_{\rho}(x)F_{\mu\nu}(x+a\hat{\rho})
U^{\dagger}_{\rho}(x) - F_{\mu\nu}(x) \nonumber\\
a \Delta^{-}_{\rho}F_{\mu\nu}(x) &=& F_{\mu\nu}(x) - U^{\dagger}_{\rho}
(x-a\hat{\rho})F_{\mu\nu}(x)U_{\rho}(x-a\hat{\rho}) \nonumber \\
g\tilde{F}_{\mu\nu}(x) &=& gF_{\mu\nu}(x) - {a^4\over6}\, \left[\Delta^+_{\mu}
\Delta^-_{\mu} + \Delta^+_{\nu}\Delta^-_{\nu} \right]\, gF_{\mu\nu}(x)
\nonumber \\
\label{impgauge}
\end{eqnarray}
The chromo-electric $\tilde{E}$ and chromo-magnetic $\tilde{B}$ fields in
$\delta H^{(3)}$ and $\delta H^{(4)}$ of Equation (\ref{ham}) are 
thus $\mathcal{O}(a^4)$ improved.

\subsection{HISQ charm and lighter quarks}

For the lighter quarks -- charm, strange and up/down -- relativistic
HISQ action \cite{hisq} is used. Apart from anything else, from practical
point of coding the bottom-light operators ($lbb, llb, l_1 l_2 b$), using
the same relativistic action for all lighter quarks offers a great
degree of simplification. The HISQ action is given by,
\begin{equation}
\mathcal{S} = \sum\limits_x \overline{l}(x)\, (\gamma^{\mu}
D_{\mu}^{\text{HISQ}} +m)\,l(x) \label{hisqlag}
\end{equation}
where,
\begin{equation}
D_{\mu}^{\text{HISQ}} = \Delta_{\mu}(W) -\frac{a^2}{6}\, (1+\epsilon)
\, \Delta_{\mu}^3(X) \label{hisq}
\end{equation}
with $W_{\mu}(x)=F_{\mu}^{\text{HISQ}}\, U_{\mu}(x)$ and $X_{\mu}(x)
= \mathcal{U}F_{\mu}U_{\mu}(x)$. The $F_{\mu}^{\text{HISQ}}$ has the
form
\begin{equation}
F_{\mu}^{\text{HISQ}}= \left(F_{\mu} -\sum_{\rho\neq\mu} {a^2(
\delta_{\rho})^2\over 2} \right)\, \mathcal{U}F_{\mu}
\label{fhisq}
\end{equation}
Here the $\mathcal{U}$ is the unitarizing operator, it unitarizes whatever
it acts on and the smearing operator $F_{\mu}$ is given by
\begin{equation}
F_{\mu} = \prod_{\rho\neq\mu} \left( 1+{a^2\delta_{\rho}^{(2)}\over 4}
\right). \label{fmu} 
\end{equation}
The $\delta_{\rho}$ and $\delta_{\rho}^{(2)}$ in the Equations (\ref{fhisq})
and (\ref{fmu}) are covariant first and second order derivatives.
Because HISQ action reduces $\mathcal{O}(\alpha_s a^2)$ discretization
errors found in Asqtad action, it is well suited for $s$ and
$u/d$ quarks. The parameter $\epsilon$ in the coefficient of Naik
term can be appropriately tuned to use the action for $c$ quark. For
$s$ and $u/d$ quarks, the $\epsilon = 0$. Later, in the Table
\ref{tab:latt_mbcs} we listed the parameters used for HISQ quarks. We
have taken the values of $\epsilon$ from \cite{eric} and used MILC
subroutines for generating HISQ propagators.

Since HISQ action is diagonal in spin space, propagators obtained do not have
any spin structure. The full $4 \times 4$ spin structure can be regained
by multiplying the propagators by Kawamoto-Smit multiplicative phase
factor \cite{kawamoto},
\begin{equation}
\Omega(x) = \prod_{\mu=1}^4(\gamma_{\mu})^{x_{\mu}} = \gamma_{1}^{x_1}
\gamma_{2}^{x_2} \gamma_{3}^{x_3} \gamma_{4}^{x_4}.
\label{staggop}
\end{equation}
MILC library uses a different representation of $\gamma$ matrices than
the ones used in NRQCD. However, $\gamma$ matrices of these two
representations are related by the unitary transformation of the form
\begin{equation}
S \, \gamma^{\text{MILC}}_{\mu}\, S^{\dagger} = \gamma^{\text{NR}}_{\mu}
\hspace{0.1in} \text{where,} \hspace{0.1in} S = {1\over \sqrt{2}}
\begin{pmatrix}
  \sigma_y && \sigma_y\\
  -\sigma_y && \sigma_y
   \end{pmatrix}\label{smatrix} 
\end{equation}

\section{TWO-POINT FUNCTIONS} \label{sec_operator}

In this section we discuss the construction of the bottom baryons by
combining spin and color indices of the appropriate quark fields to
form necessary baryon operators and two-point functions. The
$b$ quark field is universally represented with $Q$ throughout the
paper. It is defined later in the Equation (\ref{b_field}).

\subsection{Bottom-bottom meson two-point function}

After the $b$ quark propagators are generated according to
Equation (\ref{green}),
we calculate the masses of bottomonium states from the exponential
fall-off of two-point functions {\em i.e.} correlators of the state
with quantum numbers of interest. The meson creation operators are
constructed from two component quark and anti-quark creation operators
$\psi_h^{\dagger}$ and $\chi_h^{\dagger}$ \cite{davies1,davies2}. As
antiquarks transform as $\bar3$ under color rotation, we rename the
antiquark spinor as $\chi_h \equiv \chi_h^*$ \cite{thacker}.
The meson creation operator is thus,
\begin{equation}
\mathcal{O}_{hh}(x) = \psi_h^{\dagger}(x)\, \Gamma \,\chi_h(x).
\label{hhintpol}
\end{equation}
Heavy-heavy {\em i.e.} bottom-bottom meson two-point function is then
given by \cite{thacker,trottier},
\begin{eqnarray}
 C_{hh}(\vec{p},t) &=& \sum_{\vec{x}} \langle \mathcal{O}^\dagger_{hh}(x)
\,\mathcal{O}_{hh}(0) \rangle \nonumber \\
 &=& \sum_{\vec{x}}e^{i\vec{p} \cdot\vec{x}}\, \text{Tr}\, \left[
G^{\dagger}(x,0)\,\Gamma^{\dagger}_{\text{sink}}\, G(x,0)\,
\Gamma_{\text{src}} \right] \nonumber \\
 &&  \label{hhmeson}
\end{eqnarray}
$\Gamma_{\text{sink}}=\Gamma_{\text{src}}=I$ and $\sigma_i$ for the
pseudoscalar and vector mesons respectively. Heavy-heavy propagator
$G(x,0)$ is a $2 \times 2$ matrix in spin space. If we think $G(x,0)$
as a $4 \times 4$ matrix with vanishing lower components then we can
rewrite the above Equation (\ref{hhmeson}) as \cite{meinel}
\begin{equation}
C_{hh}(\vec{p},t) = \sum_{\vec{x}}e^{i\vec{p} \cdot \vec{x}} \,
\text{Tr} \,\left[ \gamma_5\, G^{\dagger}(x,0)\, \gamma_5\,
\Gamma^{\dagger}_{\text{sink}}\, G(x,0)\, \Gamma_{\text{src}} \right] 
\label{llmeson}
\end{equation}
where $\Gamma$ matrices now changed to $\Gamma_{\text{sink}} =
\Gamma_{\text{src}} = \gamma_5$ and $\gamma_i$ for pseudoscalar and
vector mesons respectively.
In Equation (\ref{llmeson}) we have used the non-relativistic Dirac
representation of $\gamma$ matrices. In the Equations (\ref{hhmeson}),
(\ref{llmeson}) and (\ref{hlmeson}) the trace is taken over both the
spin and color indices.

\subsection{Heavy-light meson two-point function}

As discussed above, $b$ quark field $\psi_h$ has only two spin components.
We convert it to a 4-component spinor having vanishing lower components
\begin{equation}
Q=\begin{pmatrix} \psi_h \\ 0 \end{pmatrix} \label{b_field}
\end{equation}
This helps us to combine the $b$ and light quark fields in the usual way,
\begin{equation}
\mathcal{O}_{hl}(x) = \bar{Q}(x)\, \Gamma\, l(x) \label{hlintpol}
\end{equation}
where $l(x)$ stands for the light quark fields, $\bar{Q} = Q^{\dagger}
\gamma_4$ and depending on pseudoscalar and vector mesons $\Gamma =
\gamma_5$ and $\gamma_i$ respectively. Note that in the Dirac {\em i.e.}
NR representation of $\gamma$-matrices $\gamma_4 \,Q = Q$. The zero
momentum bottom-light two-point function becomes \cite{burch,wingate},
\begin{eqnarray}
C_{hl}(t) &=& \sum_{\vec{x}} \langle \mathcal{O}^{\dagger}_{hl}(x)\,
\mathcal{O}_{hl}(0) \rangle \nonumber\\
 &=& \sum_{\vec{x}} \text{Tr}\, \left[\gamma_5 \,M^{\dagger}(x,0)\,
\gamma_5 \, \Gamma^{\dagger}_{\text{sink}} G(x,0)\, \Gamma_{\text{src}}
 \right] \label{hlmeson}
\end{eqnarray}
where $M(x,0)$ is the light quark propagator. It has the usual full
$4 \times 4$ spin structure. As before, $G(x,0)$ is the $b$ quark
propagator having vanishing lower components. However, before
implementing Equation (\ref{hlmeson}), $G(x,0)$ has to be rotated to
the MILC basis.

\subsection{Bottom baryon two-point functions} \label{barop}

The bottom quark field $Q$ has vanishing lower components and hence can
be projected to positive parity states only. Besides, the use of $\Gamma
= C\gamma_5$ in a diquark operator made from same flavor {\em i.e.}
$l^T \,C\gamma_5\,l$ is not allowed by Pauli exclusion principle.
In other words, the insertion of $C\gamma_5$ between two quark fields
of same flavor creates a combination which is antisymmetric in spin
indices, while the presence of $\epsilon_{abc}$ makes the combination
antisymmetric in color indices. This makes the overall operator become
symmetric under the interchange of the same flavored quark fields. Keeping
this in mind, the constructions of various bottom baryon two-point
functions are described below. \newline

\noindent
\textbf{Triply bottom baryon}: Triply bottom baryon operator is defined
by
\begin{equation}
\left( \mathcal{O}^{hhh}_k \right)_\alpha = \epsilon_{abc} \, \left(
{Q^a}^T C\gamma_k \, Q^b \right) \, Q^c_\alpha \label{hhhintpol}
\end{equation}
where $C=\gamma_4\gamma_2$. Here $a,b,c$ are the color indices, $\alpha$
is the spinor index and $k$ is the Lorentz index which runs from 1 to 3.
The zero momentum two-point function reads \cite{meinel}
\begin{eqnarray}
\lefteqn{\hspace{-0.2in} C_{jk;\alpha\delta}^{hhh}(t) } \nonumber \\
 &=&  \sum_{\vec{x}} \left\langle [\mathcal{O}^{hhh}_j(x)]_\alpha \,
[\mathcal{O}^{hhh}_k (0)]^\dagger_\delta \right\rangle \nonumber \\
 &=&\sum_{\vec{x}} \epsilon_{abc} \epsilon_{fgh}\, G^{ch}_{\alpha\delta}
(x,0) \times \nonumber \\
 &&  \hspace{0.23in} \text{Tr}\, \left[ C\gamma_j \, G^{bg}(x,0)\,
\gamma_k\gamma_2 \,{G^{af}}^T(x,0)\right] \label{hhh_cor}
\end{eqnarray}
In the above Equation (\ref{hhh_cor}) and the subsequent ones, the
transpose and traces are taken over spin indices. Baryon operators
having $C\gamma_k$ in the diquark component have overlap with both 
spin-${3\over2}$ and ${1\over 2}$ states. For example, correlator 
defined in Equation (\ref{hhh_cor}) can be written explicitly as
an overlap with both spin-${3\over2}$ and ${1\over 2}$ states
\cite{bowler},
\begin{equation}
C^{hhh}_{ij}(t) = Z^2_{3/2}\, e^{-E_{3/2}t}\Pi\, P^{3/2}_{ij} +
Z^2_{1/2}\, e^{-E_{1/2}t}\Pi\, P^{1/2}_{ij}
\end{equation}
where $\Pi=(1+\gamma_4)/2$ and the spin projection operators $P^{3/2}_{ij}
=\delta_{ij} -\gamma_i \gamma_j/3$ and $P^{1/2}_{ij}=\gamma_i \gamma_j/3$.
The individual contribution to the respective spin states can be obtained
by taking appropriate projections,
\begin{eqnarray}
P^{3/2}_{ij} \; C^{hhh}_{jk} &=& Z^2_{3/2}\Pi e^{-E_{3/2}t}\,P^{3/2}_{ik}
\nonumber\\
P^{1/2}_{ij} \; C^{hhh}_{jk} &=& Z^2_{1/2}\Pi e^{-E_{1/2}t}\, P^{1/2}_{ik}
\label{bar_sep}
\end{eqnarray}
In this paper, we use these projections to separate the different spin
states. We would like to point out that the spin-${1\over 2}$ state of
triply bottom baryon is not a physical state as it violates Pauli
exclusion principle even though we can take the projection in practice.
\newline

\noindent
\textbf{Bottom-bottom-light baryon}: Interpolating operator for baryons,
having two $b$ quarks and a light quark, can be constructed in two ways
based on how the diquark component is formed \cite{flynn}.
\begin{eqnarray}
(\mathcal{O}^{hhl}_k)_\alpha &=& \epsilon_{abc}\, \left({Q^a}^T C\gamma_k
Q^b \right)\, l^c_\alpha \label{hhdq} \\
(\mathcal{O}^{hlh}_k)_\alpha &=& \epsilon_{abc}\, \left({Q^a}^T C\gamma_k
l^b \right)\, Q^c_\alpha \label{hldq}
\end{eqnarray}
The corresponding baryon correlators are
\begin{eqnarray}
\lefteqn{\hspace{-0.2in} C^{hhl}_{jk;\alpha\delta}(t) } \nonumber \\
 &=& \sum_{\vec{x}} \left\langle [\mathcal{O}^{hhl}_j(x)]_\alpha\,
[\mathcal{O}^{hhl}_k(0)]^\dagger_\delta \right\rangle \nonumber \\
 &=& \sum_{\vec{x}} \epsilon_{abc} \epsilon_{fgh} \left[M^{ch}(x,0) \,
\gamma_4 \right]_{\alpha\delta} \times \nonumber \\
 && \hspace{0.23in} \text{Tr} \, \left[\gamma_4\gamma_2\gamma_j\, G^{bg}
(x,0)\, \gamma_k\gamma_2 \,{G^{af}}^T(x,0) \right]\label{hhl_cor} \\
\lefteqn{\hspace{-0.2in} C^{hlh}_{jk;\alpha\delta}(t) } \nonumber \\
 &=&\sum_{\vec{x}} \left\langle [\mathcal{O}^{hlh}_j(x)]_\alpha \,
[\mathcal{O}^{hlh}_k(0)]^\dagger_\delta \right\rangle \nonumber \\
 &=&\sum_{\vec{x}} \epsilon_{abc} \epsilon_{fgh} \,G^{ch}_{\alpha
\delta}(x,0) \times \nonumber \\
 && \hspace{0.23in} \text{Tr}\, \left[\gamma_4\gamma_2\gamma_j M^{bg}
(x,0)\, \gamma_k\gamma_2 \, {G^{af}}^T(x,0) \right] \label{hlh_cor}
\end{eqnarray}
The propagator $G(x,0)$ is required to be converted to
MILC basis using the unitary matrix $S$ defined in Equation
(\ref{smatrix}).

An additional spin-${1\over 2}$ operator can be defined for the
$\mathcal{O}^{hlh}$ type operator as
\begin{equation}
(\mathcal{O}^{hlh}_5)_\alpha = \epsilon_{abc}\, ({Q^a}^T\, C\gamma_5\,
l^b)\, Q^c_\alpha \label{hlh_12}
\end{equation}
The two-point function for this operator is obtained by replacing
$\gamma_j$ and $\gamma_k$ by $\gamma_5$ in Equation (\ref{hlh_cor}).
We cannot have a $C\gamma_5$ between two $Q$ in diquark and hence no
$\mathcal{O}^{hhl}_5$.
\begin{table}[h]
\caption{Operators for triple and double bottom baryons. $Q$ is used
for $b$ field and $l$ for any of the $c$, $s$, $u/d$ lighter quarks.}
\begin{ruledtabular}
\begin{tabular}{cccc}
Baryon & Quark content & $J^P$ & Operator \\ \hline
$\Omega_{bbb}$ & $bbb$ & ${3\over2}^+, {1\over 2}^+$ & $\epsilon_{abc}
({Q^a}^T C\gamma_k Q^b)\,Q^c$   \\
$\Omega_{cbb}^\star, \Omega_{cbb}$ & $cbb$ & ${3\over2}^+, {1\over 2}^+$
& $\epsilon_{abc}({Q^a}^T C\gamma_k Q^b)\,l^c$   \\
$\widetilde{\Omega}_{cbb}^\star, \widetilde{\Omega}_{cbb}$ & $cbb$ &
${3\over2}^+, {1\over 2}^+$ & $\epsilon_{abc}({Q^a}^T C\gamma_k l^b)
\,Q^c$ \\
$\Omega_{cbb}^{\prime}$ & $cbb$ & ${1\over2}^+$ & $\epsilon_{abc}({Q^a}^T
C\gamma_5 l^b)\,Q^c$   \\
$\Omega_{bb}^\star, \Omega_{bb}$ & $sbb$ & ${3\over2}^+, {1\over 2}^+$
& $\epsilon_{abc}({Q^a}^T C\gamma_k Q^b)\,l^c$ \\
$\widetilde{\Omega}_{bb}^\star, \widetilde{\Omega}_{bb}$ & $sbb$ &
${3\over2}^+, {1\over 2}^+$ & $\epsilon_{abc}({Q^a}^T C\gamma_k l^b)
\,Q^c$ \\
$\Omega_{bb}^{\prime}$ & $sbb$ & ${1\over2}^+$ & $\epsilon_{abc}({Q^a}^T
C\gamma_5 l^b)\, Q^c$   \\
$\Xi_{bb}^\star, \Xi_{bb}$ & $ubb$ & ${3\over2}^+, {1\over 2}^+$ &
$\epsilon_{abc}({Q^a}^T C\gamma_k Q^b)\,l^c$   \\
$\widetilde{\Xi}_{bb}^\star, \widetilde{\Xi}_{bb}$ & $ubb$ & ${3\over2}^+,
{1\over 2}^+$ & $\epsilon_{abc}({Q^a}^T C\gamma_k l^b)\,Q^c$   \\
$\Xi_{bb}^{\prime}$ & $ubb$ & ${1\over2}^+$ & $\epsilon_{abc}({Q^a}^T
C\gamma_5 l^b) \,Q^c$   \\
\end{tabular}
\end{ruledtabular}
\label{tab:multi_b_barop}
\end{table}

In Table \ref{tab:multi_b_barop} we tabulated the full list of triple
and double bottom baryon operators that are used in this work. We have
broadly followed the nomenclature adopted in \cite{brown} but with
certain modifications as needed for this work. The baryons having the
same quark content and $J^P$ are obtained in two different ways, as
mentioned above. The operators with ``tilde", for instance
$\widetilde{\Omega}_{bb} \; (1/2^+)$, are obtained by projecting the
relevant $(Q\,C\gamma_k\,l)\,Q$ operator with $P_{ij}^{1/2}$. The
operators with ``prime", such as $\Omega_{bb}^\prime \;(1/2^+)$,
are obtained from $(Q\,C\gamma_5\,l)\,Q$ diquark construction.
The ``prime" states so constructed on lattice correspond to the
``prime" continuum states, such as $\Xi_b^\prime$ or till unobserved
$\Omega_{cb}^\prime$ etc. It is obvious that baryon states calculated
by projecting out definite spin states from a two-point function share
the same interpolating operator. The star-ed baryons are for $J^P =
3/2^+$ states. \newline

\noindent
\textbf{Bottom-light-light baryon}: The natural choice for interpolating
operator, as motivated by Heavy Quark Effective Theory \cite{hqet}, in
$llh$-baryon kind is
\begin{equation}
(\mathcal{O}^{h l_1 l_2}_k)_\alpha = \epsilon_{abc} \, ({l_1^a}^T C
\gamma_k l_2^b) \, Q^c_\alpha \label{hll_hqet}
\end{equation}
and the corresponding two-point function is
\begin{eqnarray}
\lefteqn{\hspace{-0.2in} C^{h l_1 l_2}_{jk;\alpha\delta}(t) } \nonumber \\
 &=& \sum_{\vec{x}} \left\langle [\mathcal{O}^{h l_1 l_2}_j(x)]_\alpha
\, [\mathcal{O}^{h l_1 l_2}_k (0)]^\dagger_\delta \right\rangle\nonumber \\
 &=& \sum_{\vec{x}} \epsilon_{abc} \epsilon_{fgh}\, G^{ch}_{\alpha\delta}
(x,0) \times \nonumber \\
 && \hspace{0.23in} \text{Tr}\, \left[\gamma_4\gamma_2\gamma_j \, M_2^{bg}
(x,0) \, \gamma_k\gamma_2\gamma_4 \, {M_1^{af}}^T(x,0) \right] \label{llh_cor}
\end{eqnarray}
In HISQ formalism for light quarks, the corresponding propagators $M_1(x,0)$
and $M_2(x,0)$ have the same Kawamoto-Smit multiplicative factor $\Omega(x)$.
Hence they have the same spin structure irrespective of color indices. As
a result, the trace over spin indices in Equation (\ref{llh_cor}) vanishes
if $\gamma_j\neq\gamma_k$. Therefore, we can not separate the two
spin-${3\over2}$ and ${1\over 2}$ states. If we want to use the same diquark
structure as in Equation (\ref{hll_hqet}), for baryons having different
light quark flavors we can define the spin-${1\over2}$ operator by
\begin{equation}
(\mathcal{O}^{h l_1 l_2}_5)_\alpha = \epsilon_{abc}\, ({l_1^a}^T \,C
\gamma_5 \, l_2^b) \, Q^c_\alpha \label{llh_5}
\end{equation}
and the corresponding two-point function is
\begin{eqnarray}
\lefteqn{\hspace{-0.2in} C^{h l_1 l_2}_{55;\alpha\delta}(t) } \nonumber \\
 &=& \sum_{\vec{x}} \left\langle [\mathcal{O}^{h l_1 l_2}_5 (x)]_\alpha
\, [\mathcal{O}^{h l_1 l_2}_5 (0)]^\dagger_\delta \right\rangle\nonumber \\
 &=& \sum_{\vec{x}} \epsilon_{abc} \epsilon_{fgh}\, G^{ch}_{\alpha\delta}
(x,0) \times \nonumber \\
 && \hspace{0.23in} \text{Tr}\, \left[\gamma_4\gamma_2\gamma_5 \, M_2^{bg}
(x,0) \, \gamma_5\gamma_2\gamma_4 \, {M_1^{af}}^T(x,0) \right]
\label{llh5_cor}
\end{eqnarray}
Besides, instead of Equation (\ref{hll_hqet}), we can choose our
($h l_1 l_2$)-operator as
\begin{equation}
 (\mathcal{O}^{h l_2 l_1}_k)_\alpha = \epsilon_{abc}\, ({Q^a}^T C
\gamma_k \, l_2^b) \, l^c_{1\alpha} \label{hllintpol}
\end{equation}
The two-point function is now
\begin{eqnarray}
C^{h l_2 l_1}_{jk,\alpha\delta}(t) &=&
\sum_{\vec{x}} \epsilon_{abc} \epsilon_{fgh} \left[M_1^{ch}(x,0)\,
\gamma_4 \right]_{\alpha\delta} \times \nonumber \\
 && \text{Tr} \, \left[\gamma_4\gamma_2\gamma_j \,M_2^{bg}(x,0)\,
\gamma_k\gamma_2 \,{G^{af}}^T(x,0) \right] \nonumber \\
 && \label{hll_cor}
\end{eqnarray} 
Because the light quark propagators $M_1(x,0)$ and $M_2(x,0)$ are
proportional to each other, the relative positions of the quark 
fields $l_1$ and $l_2$ in Equation (\ref{hllintpol}) is irrelevant.
Interpolating operator defined in Equation (\ref{hllintpol}) has
overlap with both spin-${3\over2}$ and ${1\over 2}$ states and
can be projected out by appropriate projection operators $P_{ij}^{1/2,
\;3/2}$.

As before, we can also define an additional spin-${1\over 2}$ operator
here too,
\begin{equation}
(\mathcal{O}^{h l_2 l_1}_5)_\alpha = \epsilon_{abc}\, ({Q^a}^T C\gamma_5
\, l_2^b)\, l^c_{1\alpha}
\end{equation}
The two-point function for this operator has the same form as in Equation
(\ref{hll_cor}) with $\gamma_j$ and $\gamma_k$ replaced by $\gamma_5$.
\begin{table}[h]
\caption{Operators for single bottom baryons. $Q$ is used for
$b$ field as before. Interchange in the position of two lighter quarks
keeps the operator unchanged.}
\begin{ruledtabular}
\begin{tabular}{cccc}
Baryon & Quark content & $J^P$ & Operator \\ \hline
$\widetilde{\Omega}_{ccb}^\star, \widetilde{\Omega}_{ccb}$ & $ccb$ &
${3\over2}^+, {1\over 2}^+$ & $\epsilon_{abc}({Q^a}^T C\gamma_k c^b)
\,c^c$ \\
$\Omega_{ccb}^{\prime}$ & $ccb$ & ${1\over2}^+$ & $\epsilon_{abc}
({Q^a}^T C\gamma_5 c^b)\,c^c$   \\
$\Omega_{cb}$ & $scb$ & ${1\over2}^+$ & $\epsilon_{abc}({s^a}^T
C\gamma_5 c^b)\,Q^c$   \\
$\widetilde{\Omega}_{cb}^\star, \widetilde{\Omega}_{cb}$ & $scb$ &
${3\over2}^+, {1\over 2}^+$ & $\epsilon_{abc}({Q^a}^T C\gamma_k c^b)
\,s^c$   \\
$\Omega_{cb}^\prime$ & $scb$ & ${1\over2}^+$ & $\epsilon_{abc}({Q^a}^T
C\gamma_5 c^b)\,s^c$   \\
$\Xi_{cb}$ & $ucb$ & ${1\over2}^+$ & $\epsilon_{abc}({u^a}^T C\gamma_5
c^b) \,Q^c$   \\
$\widetilde{\Xi}_{cb}^\star, \widetilde{\Xi}_{cb}$ & $ucb$ & ${3\over2}^+,
{1\over 2}^+$ & $\epsilon_{abc}({Q^a}^T C\gamma_k c^b) \,u^c$   \\
$\Xi_{cb}^\prime$ & $ucb$ & ${1\over2}^+$ & $\epsilon_{abc}({Q^a}^T
C\gamma_5 c^b) \,u^c$  \\
$\widetilde{\Omega}_{b}^\star, \widetilde{\Omega}_b$ & $ssb$ &
${3\over2}^+, {1\over 2}^+$ & $\epsilon_{abc}({Q^a}^T C\gamma_k s^b)
\,s^c$ \\
$\Omega_{b}^{\prime}$ & $ssb$ & ${1\over2}^+$ & $\epsilon_{abc}({Q^a}^T
C\gamma_5 s^b)\,s^c$   \\
$\Xi_{b}$ & $usb$ & ${1\over2}^+$ & $\epsilon_{abc}({u^a}^T C\gamma_5
s^b)\,Q^c$   \\
$\widetilde{\Xi}_{b}^\star, \widetilde{\Xi}_b$ & $usb$ & ${3\over2}^+,
{1\over 2}^+$ & $\epsilon_{abc}({Q^a}^T C\gamma_k s^b) \,u^c$   \\
$\Xi_{b}^\prime$ & $usb$ & ${1\over2}^+$ & $\epsilon_{abc}({Q^a}^T
C\gamma_5 s^b) \,u^c$   \\
$\widetilde{\Sigma}_{b}^\star, \widetilde{\Sigma}_b$ & $uub$ &
${3\over2}^+, {1\over 2}^+$ & $\epsilon_{abc}({Q^a}^T C\gamma_k u^b)
\,u^c$ \\
$\Sigma_{b}^{\prime}$ & $uub$ & ${1\over2}^+$ & $\epsilon_{abc}({Q^a}^T
C\gamma_5 u^b)\, u^c$   \\
$\Lambda_{b}$ & $udb$ & ${1\over2}^+$ & $\epsilon_{abc}({u^a}^T
C\gamma_5 d^b)\,Q^c$   \\
\end{tabular}
\end{ruledtabular}
\label{tab:one_b_barop}
\end{table}

In Table \ref{tab:one_b_barop} we tabulate our full list of single
bottom baryon operators that we made use of in this work. The ``tilde"
and ``prime" states that appear in the table have been explained before
in the context of multi bottom baryon operators. \newline

\noindent
\textbf{Light baryon}: We occasionally need charmed baryon states like
$qcc$ or $qqc$, where $q$ is any of $s$ or $u/d$ quarks or both, hence
we include a discussion on charmed baryon operators. The $c$-quark in
the present case is relativistic. For the reason discussed above, with HISQ
action for lighter quarks we can define only the spin-${1\over2}$
operators. Consider a $(l_1 l_2 l_3)$-baryon where at least two quarks
are differently flavored, say $l_1 \ne l_2$. The spin-${1\over 2}$
operator and the corresponding two-point function in such case is 
\begin{eqnarray}
(\mathcal{O}^{l_1 l_2 l_3}_5)_\alpha &=& \epsilon_{abc}\, ({l_1^a}^T
C\gamma_5 \, l_2^b)\, l_{3\alpha}^c \label{lllintpol} \\
C^{l_1 l_2 l_3}_{55,\alpha\delta}(t) &=&\sum_{\vec{x}} \epsilon_{abc}
\epsilon_{fgh}\, \left[M_3^{ch}(x,0)\,\gamma_4 \right]_{\alpha\delta}
\times \nonumber \\
 && \text{Tr} \,\left[\gamma_4\gamma_2\gamma_5 \,M_2^{bg}(x,0) \,
\gamma_5\gamma_2\gamma_4 \,{M^{af}}^T_1(x,0)\right] \nonumber \\
 && \label{lll_cor}
\end{eqnarray}
The two light baryon states $(J^P = 1/2^+)$ that we are interested
in this work are,
\begin{eqnarray}
\Sigma_c \; (uuc) &:& \hspace{0.4in} \epsilon_{abc}({c^a}^T C\gamma_5
u^b)\,u^c \nonumber \\
\Xi_{cc} \; (ucc) &:& \hspace{0.4in} \epsilon_{abc}({c^a}^T C\gamma_5
u^b)\,c^c \label{c_bar}
\end{eqnarray}
 
\section{Simulation details} \label{sec_simul}

We calculated the bottom baryon spectra using the publicly available
$N_f=2+1$ Asqtad gauge configurations generated by MILC collaboration.
Details about these lattices can be found in \cite{bazavov}. It uses
Symanzik-improved L\"{u}scher-Weisz action for the gluons and Asqtad 
action \cite{asqtad1, asqtad2} for the sea quarks. The lattices we
choose have a fixed ratio of $am_l/am_s=1/5$ and lattice spacings
ranging from 0.15 fm to 0.09 fm corresponding to the same physical
volume. We have not determined the lattice spacings independently but
use those given in \cite{bazavov}. In Table \ref{tab:milclatt} we
listed the ensembles used in this work. 

\begin{table}[h]
\caption{MILC configurations used in this work. The gauge coupling is
$\beta$, lattice spacing $a$, $u/d$ and $s$ sea quark masses are
$m_l$ and $m_s$ respectively and lattice size is $L^3\times T$. The
$N_{\text{cfg}}$ is number of configurations used in this work.}
\begin{ruledtabular}
\begin{tabular}{cccccc}
$\beta={10/ g^2}$ & $a$(fm) & $am_l$ & $am_s$ & $L^3\times T$ &
$N_{\text{cfg}}$    \\       
\hline
6.572 & 0.15 & 0.0097 & 0.0484 & $16^3\times 48$ & 400 \\
6.76 & 0.12 & 0.01 & 0.05 & $20^3\times 64$ & 400 \\
7.09 & 0.09 & 0.0062 & 0.031 & $28^3\times 96$ & 300 \\
\end{tabular}
\end{ruledtabular}
\label{tab:milclatt}
\end{table}

In NRQCD the rest mass term does not appear in Equation (\ref{ham}) and
therefore we cannot determine hadron masses from their energies at
zero momentum directly from the exponential fall-off of the correlation
functions. Instead, we calculate the kinetic mass $M_k$ of heavy-heavy
mesons from its energy-momentum relation, which to $\mathcal{O}(p^2)$ 
is \cite{fermilab},
\begin{eqnarray}
&& E(p) = E(0) + \sqrt{p^2 + M_k^2} - M_k \nonumber \\
\Rightarrow \;\;\; && E(p)^2 = E(0)^2 + {E(0)\over M_k}\,p^2 \label{kin-mass}
\end{eqnarray} 
We calculate the $E(p)$ at different values of lattice momenta $p=2n\pi/L$
where, $n =$ (0,0,0), (1,0,0), (1,1,0), (1,1,1), (2,0,0), (2,1,0) and (2,1,1).
\newline

\noindent
\textbf{$\mathbf{m_b}$ tuning}: The $b$ quark mass is tuned from the
spin average $\Upsilon$ and $\eta_b$ masses
\begin{equation}
M_{b\bar{b}} = {3M_{\Upsilon} + M_{\eta_b}\over 4} \label{mbspinavg}
\end{equation}
using kinetic mass for both $\Upsilon$ and $\eta_b$. The experimental
value to which $M_{b\bar{b}}$ is tuned to is not 9443 MeV, as obtained
from spin averaging $\Upsilon$ (9460 MeV) and $\eta_b$ (9391 MeV)
experimental masses, but to an appropriately adjusted value of 9450
MeV \cite{eric}, which we denote as $M_{\text{phys}}^{\text{mod}}$
later in the Equation (\ref{mass_formula}). The reasons being, firstly
electromagnetic interaction among the quarks are not considered here.
Secondly, the disconnected diagrams while computing two-point function
are also not considered thus not allowing $b$, $\bar{b}$ quarks to
annihilate to gluons. And finally, we do not have sea $c$ quarks in
our simulation. For a more detailed discussion on this, see \cite{eric}.

The $b$ quark mass $m_b$ and the coefficient $c_4$ in Equation
(\ref{ham}) are then tuned to obtain modified spin average mass and
the hyperfine splitting of $\Upsilon$ and $\eta_b$, which is $\sim
60 - 65$ MeV \cite{zerf}. In order to achieve the desired hyperfine
splittings, we tuned only $c_4$ since at $\mathcal{O}(1/m_b)$ this is
the only term that contains Pauli spin matrices. Therefore, it allows
the mixing of spin components of $\psi_h$. This term contributes
maximally to the
hyperfine splitting compared to the others that contain Pauli spin
matrices. The one-loop radiative correction to $c_4$ \cite{c4rad} has
been used to tune hyperfine splitting in \cite{c4hf} where it was found
to change, but only mildly, over lattice spacings $\sim 0.15 - 0.09$ fm
for $N_f = 2 + 1 + 1$ HISQ gauge configurations. In our present mixed
action study, the changes in the tuned $c_4$ on various lattice ensembles
are small enough. Taking an average of those values we choose $c_4
= 1.9$ for which the hyperfine splittings obtained on three different
lattices 0.15, 0.12 and 0.09 fm are 60.6 MeV, 61.1 MeV and 61.8 MeV
respectively.

All other coefficients $c_i$ in Equation (\ref{ham}) are set to 1.0.
We set stability factor $n=4$ throughout our simulation. The Table
\ref{tab:latt_mbcs} lists the values of $m_b$ used in this work.
\begin{table}[h]
\caption{Tuned $b$, $c$ and $s$ quark bare masses for lattices used
in this work. For $s$-quark mass, we mentioned the particle states
to which it is tuned to. The values of $\epsilon$-parameter used for 
$c$-quark are given in the last column.}
\begin{ruledtabular}
\begin{tabular}{ccclll}
$a$ & $am_b$ & $am_c$ & $am_s$ & $am_s$ & $\epsilon$ \\       
(fm) & & & ($\eta_s$) & ($B_s$) & \cite{eric} \\
  \hline
0.15 & 2.76 & 0.850 & 0.065  & 0.215  & -0.34 \\
0.12 & 2.08 & 0.632 & 0.049  & 0.155  & -0.21 \\
0.09 & 1.20 & 0.452 & 0.0385 & 0.114  & -0.115 \\
\end{tabular}
\end{ruledtabular}
\label{tab:latt_mbcs}
\end{table}

\noindent
\textbf{$\mathbf{m_c}$ tuning}: The $c$-quark mass is tuned pretty much
in the same way as $m_b$, except that $M_{c\bar{c}}$ is tuned to the spin
average of $J/\psi$ and $\eta_c$ experimental masses. In this case,
however, the adjustment to spin averaged value due to the absence of
electromagnetic interaction, $c$-quarks in sea and disconnected diagrams
are very small and hence neglected. The bare $c$-quark masses used
in this work are given in Table \ref{tab:latt_mbcs}.
\newline

\noindent
\textbf{$\mathbf{m_s}$ tuning}: The $s$-quark mass is tuned to two
different values. In the first case, we tune to the mass of fictitious
$s\bar{s}$ pseudoscalar meson $\eta_s$ while in the second case to the
mass of $B_s$. The $\eta_s$ is a fictitious meson that is not allowed
to decay through $s\bar{s}$ annihilation. Hence no disconnected
diagrams arise in the two-point function calculation. From chiral
perturbation theory, its mass is estimated to be $m_{\eta_s} =
\sqrt{2m^2_K-m^2_{\pi}} =$ 689 MeV \cite{etas1,etas2}. The $s$-quark
mass thus tuned is checked against $D_s$ meson, making use of the
$c$-quark mass obtained above and found to agree with the experimental
$D_s$ (1968 MeV).
\begin{table}[h]
\caption{$D$ and $B$ meson masses in MeV with the tuned $am_b, am_c$
and $am_s$.}
\begin{ruledtabular}
\begin{tabular}{c|l|ll}
$L^3\times T$ & $B_c$ & \multicolumn{2}{c}{$D_s$} \\ \cline{3-4}
 & & $\eta_s$ & $B_s$ \\ \hline
$16^3 \times 48$ & 6260(8) & 1994(3) & 2197(2) \\
$20^3 \times 64$ & 6263(12) & 1977(4) & 2172(2) \\
$28^3 \times 96$ & 6255(10) & 1971(3) & 2167(2) \\
PDG \cite{pdg} & 6275 & 1968 & \\
\end{tabular}
\end{ruledtabular}
\label{tab:bd_meson}
\end{table}

Next we explore, tuning $m_s$ when $s$-quark is in a bound state with a
heavy $b$ quark. Here we are assuming that the potential experienced by
the $s$ quark in the field of $b$ quark in $B_s$ meson remains the same
in other strange bottom baryons and there is no spin-spin interactions
taking place between the quarks. In the infinite mass limit, the HQET
Lagrangian becomes invariant under arbitrary spin rotations \cite{hqet}.
Thereby, we can argue that for $sbb$ and $scb$ systems the spin-spin
interactions do not contribute significantly in spectrum calculation.
However, this argument is perhaps not valid in systems like $bss$
or $bsd$ but still with $s$ quark thus tuned, we possibly can obtain
their masses close to their physical masses without resorting to any
extrapolation.

\begin{figure}[h]
\scalebox{0.75}{\input{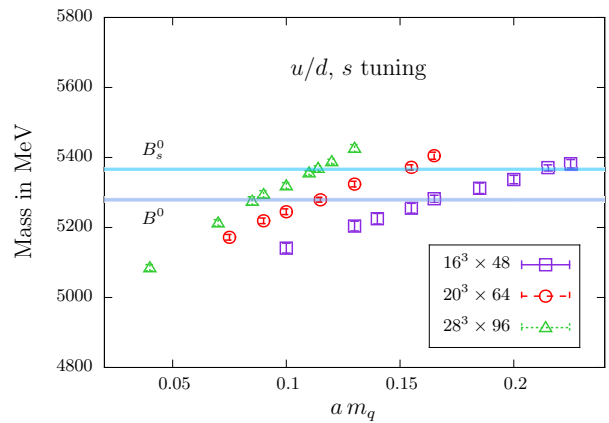}}
\caption{Tuning of $s$ and $u/d$ quark masses in various lattices. The
experimental values of $B^0$ and $B^0_s$ are shown by bands whose thickness
are to enhance visibility and have nothing to do with experimental errors.} 
\label{qtune}
\end{figure}

In this paper, we will present our results obtained at these two
different values of $m_s$. In Table (\ref{tab:latt_mbcs}) we listed
these values of $s$-quark masses. In the Table (\ref{tab:bd_meson})
we calculate $B_c$ and $D_s$ mesons using tuned $b$, $c$ and $s$
masses. As is seen, when $m_s$ is tuned with $\eta_s$ the $D_s$ mass
obtained is fairly close to PDG value whereas when tuned to $B_s$ we
see an upward shift by an average 200 MeV. We have observed similar
differences when $s$-quark appears together with $c$ in ($scb$)-baryon
masses.
\newline

\noindent
\textbf{$\mathbf{u/d}$ quarks}: For the valence $u/d$ quark mass, we
used a range of bare quark masses
varying from the lightest sea quark masses all the way to a little
above where $s$ mass is tuned to $B_s$. Whenever the mass of a bottom
baryon containing $u/d$ quark(s) is quoted, it will correspond to
$u/d$ quark mass tuned at $B$ mass. Since we are not including either
electromagnetic or isospin breaking in our calculation, we do not
distinguish between $u$ and $d$ quarks and it is always $a m_u =
a m_d$. This tuning of $u/d$ mass to $B$ works well in capturing the
$b$-baryon states containing single $u/d$, such as ($usb$) or ($ucb$) 
baryons, when compared to either PDG or other works.

In Table (\ref{tab:latt_mud}), we listed $u/d$ quark masses
($am_q$) against the lattice spacing. We show in the Figure
\ref{qtune} our strategy used to tune $m_s$ and $m_{u/d}$.
\begin{table}[h]
\caption{Values of $am_{u/d}$ used in this work. }
\begin{ruledtabular}
\begin{tabular}{lc}
$L^3\times T$ &  $am_{u/d}$ \\ \hline
$16^3\times 48$ & 0.065, 0.10, 0.13, 0.14, 0.155 \\
(0.15 fm) & 0.165, 0.185, 0.215, 0.225 \\ \hline
$20^3 \times 64$ & 0.05, 0.075, 0.090, 0.10, \\
(0.12 fm) & 0.115, 0.13, 0.155, 0.165  \\ \hline
$28^3 \times 96$ & 0.04, 0.07, 0.085, 0.09, \\
(0.09 fm) & 0.10, 0.114, 0.12, 0.13 \\
\end{tabular}
\end{ruledtabular}
\label{tab:latt_mud}
\end{table}

\noindent
The tuned $am_{u/d}$ for different lattices to use with $b$ quarks
are,
\[
16^3\times 48:\;\; 0.165 \hspace{0.2in} 20^3\times 64: 0.115
\hspace{0.2in} 28^3 \times 96: 0.085
\]
The $s$ and $u/d$ masses so tuned, $m_s/m_{u/d}$ turns out to be
$\sim 1.3$ compared to $\sim 6$ that we get when $m_s$ is obtained
from $\eta_s$ and $m_{u/d}$ is the bare sea quark mass. 

For states containing two $u/d$ quarks, such as $\Sigma_b\;(uub)$ and
$\Lambda_b\;(udb)$, we have mixed success with the above approach.
When $b$ and $u$ form diquarks $(Q\, C \gamma_{\{k,5\}} \,u)$ for
$\Sigma_b$ state, the masses obtained are consistent with other lattice
studies. However, this tuning scheme involving $B$ fails for $\Lambda_b$
where the diquark part is $(u\,C\gamma_5 \,d)$ (see the Table
\ref{tab:one_b_barop}). 
Hence for $\Lambda_b$, containing both $u$ and $d$ quarks, we have to
resort to different tuning to account for 190 MeV of mass difference with
$\Sigma_b$.  The mass of $\Lambda_b$ is obtained from this specially
tuned $m^\prime_{u/d}$ (to be used only for $\Lambda_b$). Thus tuned 
differently, the mass differences $\Xi_b -
\Lambda_b$ and $\Sigma_b^\star - \Lambda_b$ are well within $2 \sigma$
of PDG values while $\Lambda_b - B$ is about 60 MeV higher, see the
Table (\ref{tab:mass_diff}).

\section{Results and discussion} \label{sec_result}

In order to extract the masses of the bottom baryons, we perform
two-exponential uncorrelated fit to the two-point functions. We then
cross-checked it with fitting the effective masses over the same range
of time slices. However, this zero momentum energy does not directly
give us the mass of the bottom baryons because of unphysical shift in
zero of energy. To account for it, the mass is obtained considering
energy splittings,
\begin{equation}
M_{\text{latt}} = E_{\text{latt}} + {n_b\over2}\, \left(
M_{\text{phys}}^{\text{mod}} - E_{\text{latt}}^{\eta_b} \right)
\label{mass_formula}
\end{equation}
where $E_{\text{latt}}$ is the lattice zero momentum energy in MeV,
$n_b$ is the number of $b$-quarks in the bottom baryon. For bottom
mesons $(b\bar{l})$, $n_b$ is obviously always 1. As discussed before,
$M_{\text{phys}}^{\text{mod}}$ is the modified spin average mass 
of $\Upsilon$ and $\eta_b$ and is equal to 9450 MeV and $M_{\text{latt}}$
is the lattice bottom baryon mass in MeV.

In calculation of mass splittings this shift in energies is cancelled
by subtraction among energies of hadrons having equal number of $b$
quarks ($n_b$) in them. For this calculation, we use jack-knifed ratio
of the correlation functions for fitting \cite{gmo},
\begin{equation}
C^{Y-X}(t) = \frac{C^Y(t)}{C^X(t)} \;\sim \; e^{-(M_Y - M_X)\,t}
\label{splitcorr}
\end{equation}
Below in the Figure {\ref{one_b_corr_demo} we show a few correlators
for single $b$ baryons containing exclusively either two $c$ or $s$ or
$u/d$. 
\begin{figure}[h]
\scalebox{0.75}{\input{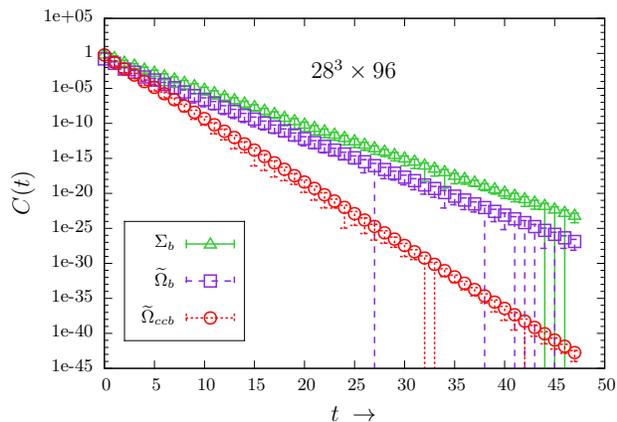}}
\caption{$\Sigma_b$, $\widetilde{\Omega}_b$ and $\widetilde{\Omega}_{ccb}$
correlators in $28^3\times 96$ lattice.} 
\label{one_b_corr_demo}
\end{figure}

The fitting range is typically chosen ($i$) looking at positions of what
we consider plateau in the effective mass plots and ($ii$) exponential
fits of the correlators. Both fittings return same masses over suitably
chosen range. In the effective mass plot Figure \ref{one_b_efm_demo},
the zero momentum energies and the errors of the same three states as
in the Figure \ref{one_b_corr_demo} are represented as bands.
\begin{figure}[h]
\scalebox{0.75}{\input{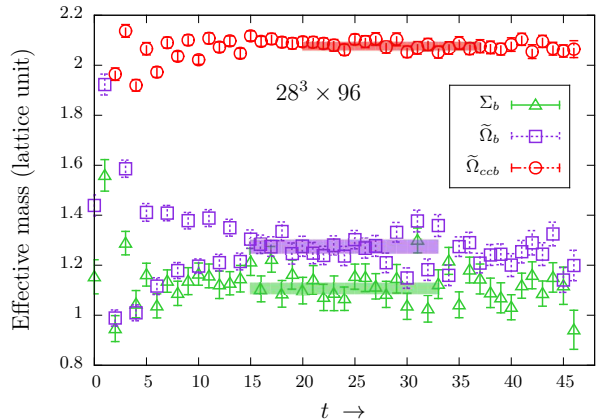}}
\caption{Effective mass plots for the states in Figure \ref{one_b_corr_demo}.
The bands are placed over what we consider plateau.}
\label{one_b_efm_demo}
\end{figure}

In these figures, we choose to present the data from $28^3 \times 96$
lattices but the data from $16^3\times 48$ and $20^3\times 64$ are
similar. Just to remind, in order to obtain the masses in MeV from
these, we need the Equation (\ref{mass_formula}). \newline

\noindent
\textbf{Single bottom baryons}: A couple of baryon states containing
one $b$ quark have been listed in the PDG \cite{pdg}, such as
$\Lambda_b \;(udb)$, $\Omega_b \;(ssb)$, $\Xi_b^\prime \;(usb)$ etc.
and they provide a good matching opportunity. In the Figure
\ref{btune_l2896}, we show the agreement of some of these baryon masses
with PDG at the tuned $m_{u/d}$ and $m_s$.

\begin{figure}[h]
\scalebox{0.75}{\input{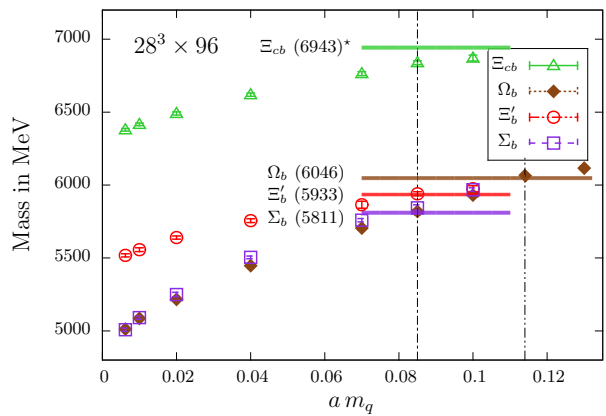}}
\caption{Variation of single $b$ baryon masses in MeV against the
same light quark masses as in Figure \ref{qtune}. $m_q = 0.085$ and
0.114 are the tuned $u/d$ and $s$ quark masses respectively, indicated
by dashed vertical lines. The bands correspond to the PDG values,
except for $\Xi_{cb}$ which is taken from \cite{brown}.}
\label{btune_l2896}
\end{figure}

In the following Tables \ref{tab:one_b_no_s}, \ref{tab:one_b_s},
\ref{tab:multi_b_no_s} and \ref{tab:multi_b_s}, we present our results
of the single and multi $b$ baryon states corresponding to the operators
given in the Tables \ref{tab:multi_b_barop} and \ref{tab:one_b_barop}.
In the columns corresponding to various lattice ensembles, we show the
masses in lattice unit, $aE_{\text{latt}}$ of the Equation
(\ref{mass_formula}). In the last column of each table, we provide the
average $M_{\text{latt}}$ and the statistical errors, calculated
assuming the 
lattice configurations of different lattice spacings are statistically 
uncorrelated. Additionally we include tables for $M_{\text{latt}}$ for
the bottom baryons for each lattice spacing in the Appendix. 

We collect our results for single bottom baryon, not containing
$s$-quark(s), in the Table \ref{tab:one_b_no_s} and those with
$s$-quark in Table \ref{tab:one_b_s}. For the $m_{u/d}$, we state
the results when the valence $m_l$ gives physical $B$ meson mass.

\begin{table}[h]
\caption{Masses, in lattice unit, of baryons involving single $b$ quark
and no $s$ quark. The bare $u/d$-quark masses are 0.165 for $16^3\times
48$, 0.115 for $20^3\times 64$ and 0.085 for $28^3\times 96$.}
\begin{ruledtabular}
\begin{tabular}{c|c|c|c|c}
Baryons & $16^3\times 48$ & $20^3\times 64$ & $28^3\times 96$ &
Average \\
 & (0.15 fm) & (0.12 fm) & (0.09 fm) & (MeV) \\ \hline
$\widetilde{\Omega}_{ccb}^\star$ & 2.954(5) & 2.497(4) & 2.088(3) &
7807(11) \\
$\widetilde{\Omega}_{ccb}$ & 2.933(5) & 2.482(3) & 2.078(3) & 
7780(9) \\
$\Omega_{ccb}^\prime$ & 2.952(4) & 2.497(3) & 2.078(3) & 
7797(11) \\
$\widetilde{\Xi}_{cb}^\star$ & 2.222(6) & 1.899(4) & 1.648(6) & 
6835(20) \\
$\Xi_{cb}$ & 2.177(11) & 1.881(4) & 1.623(5) & 6787(12) \\
$\widetilde{\Xi}_{cb}$ & 2.199(8) & 1.886(4) & 1.631(4) & 6805(16) \\
$\Xi_{cb}^\prime$ & 2.227(6) & 1.904(4) & 1.653(6) & 6843(19) \\
$\widetilde{\Sigma}_b^\star$ & 1.468(8) & 1.292(5) & 1.189(5) &
5836(22) \\
$\widetilde{\Sigma}_b$ & 1.460(7) & 1.290(3) & 1.174(6) & 5820(21) \\
$\Sigma_b^\prime$ & 1.470(7) & 1.305(3) & 1.194(9) &  5848(18) \\
$\Lambda_b$ & 1.322(7) & 1.208(6) & 1.109(9) & 5667(14) \\
\end{tabular}
\end{ruledtabular}
\label{tab:one_b_no_s}
\end{table}

Since $s$ quark has been tuned in two different ways, we quote both
the $b$-baryon masses at $\eta_s, \;B_s$ points.
\begin{table}[h]
\caption{Masses, in lattice unit, of baryons containing single $b$-quark
and $s$-quark(s).}
\begin{ruledtabular}
\begin{tabular}{c|c|c|c|c|c}
Baryons & Tuning & $16^3\times 48$ & $20^3\times 64$ & $28^3\times 96$
& Average \\
 & & (0.15 fm) & (0.12 fm) & (0.09 fm) & (MeV) \\ \hline
\multirow{2}{*}{$\widetilde{\Omega}_{cb}^\star$} & $\eta_s$ & 2.035(5)
& 1.782(5) & 1.542(3) & 6611(9) \\
	& $B_s$ & 2.292(7) & 1.957(6) & 1.693(4) & 6930(19) \\ \hline
\multirow{2}{*}{$\Omega_{cb}$} & $\eta_s$ & 2.010(8) & 1.754(5)
& 1.532(3) & 6578(9) \\
	& $B_s$ & 2.248(11) & 1.937(7) & 1.684(2) & 6893(16) \\ \hline
\multirow{2}{*}{$\widetilde{\Omega}_{cb}$} & $\eta_s$ & 2.012(7) &
1.765(5) & 1.536(3) & 6587(10) \\
	& $B_s$ & 2.267(8) & 1.943(7) & 1.686(2) & 6906(17) \\ \hline
\multirow{2}{*}{$\Omega_{cb}^\prime$} & $\eta_s$ & 2.052(5) & 1.785(5)
& 1.548(3) & 6625(8) \\
	& $B_s$ & 2.297(6) & 1.966(6) & 1.705(2) & 6946(17) \\ \hline
\multirow{2}{*}{$\widetilde{\Xi}_b^\star$} & $\eta_s$ & 0.987(4) &
0.945(2) & 0.918(3) & 5237(8) \\
	& $B_s$ & 1.541(8) & 1.352(6) & 1.235(6) & 5935(22) \\ \hline
\multirow{2}{*}{$\Xi_b$} & $\eta_s$ & 0.986(5) & 0.947(2) & 0.909(4) &
5231(11) \\
	& $B_s$ & 1.520(9) & 1.345(3) & 1.207(6) & 5901(20) \\ \hline
\multirow{2}{*}{$\widetilde{\Xi}_b$} & $\eta_s$ & 0.978(5) & 0.944(2)
& 0.904(5) & 5222(13) \\
	& $B_s$ & 1.532(11) & 1.350(4) & 1.224(4) & 5921(19) \\ \hline
\multirow{2}{*}{$\Xi_b^\prime$} & $\eta_s$ & 0.987(4) & 0.948(3)
& 0.913(5) & 5235(11) \\
	& $B_s$ & 1.544(10) & 1.366(4) & 1.238(6) & 5946(16) \\ \hline
\multirow{2}{*}{$\widetilde{\Omega}_{b}^\star$} & $\eta_s$ & 1.129(5) &
1.058(3) & 1.012(4) & 5430(11) \\
	& $B_s$ & 1.611(8) & 1.412(6) & 1.264(3) & 6019(20) \\ \hline
\multirow{2}{*}{$\widetilde{\Omega}_{b}$} & $\eta_s$ & 1.118(7) & 1.050(3)
& 0.997(4) & 5410(10) \\
	& $B_s$ & 1.600(11) & 1.411(7) & 1.264(3) & 6014(17) \\ \hline
\multirow{2}{*}{$\Omega_{b}^\prime$} & $\eta_s$ & 1.131(9) & 1.057(3)
& 1.007(2) & 5427(9) \\
	& $B_s$ & 1.615(8) & 1.425(7) & 1.295(2) & 6051(15) \\
\end{tabular}
\end{ruledtabular}
\label{tab:one_b_s}
\end{table}

As is evident from our results, the numbers coming from $s$-quark
tuned to $\eta_s$ are about 300 MeV smaller from those tuned to $B_s$
(600 MeV in baryons with two $s$). If we take $\Omega_b$ ($ssb$)
and compare with PDG value 6046 MeV, it becomes obvious.

\begin{figure}[h]
\scalebox{0.75}{\input{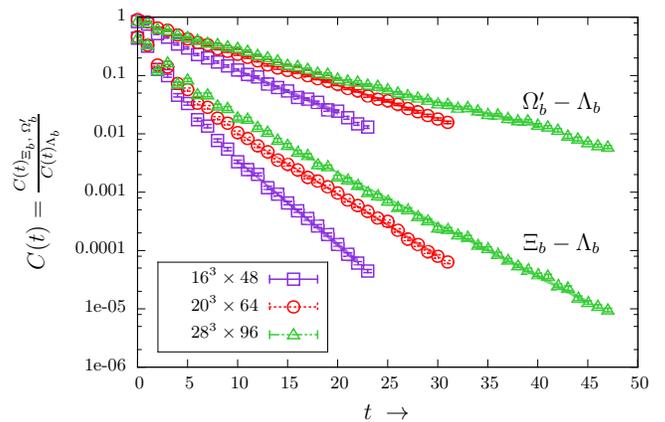}}
\caption{Ratio of correlators for the calculation of the two splittings
shown in Table \ref{tab:one_b_split}. The bands overlaid on the data
points represent single exponential fits.}
\label{corrsplitfig}
\end{figure}

Next we determine mass differences in single bottom sector
including the hyperfine splittings.

\begin{table}[h]
\caption{Single bottom baryons mass splittings in MeV.}
\begin{ruledtabular}
\begin{tabular}{c|c|c|c|c}
Baryon & $16^3 \times 48$ & $20^3\times 64$ & $28^3\times 96$ & Average \\
splittings & (MeV) & (MeV) & (MeV) & (MeV) \\ \hline
$\widetilde{\Omega}_{ccb}^\star - \widetilde{\Omega}_{ccb}$ & 28(3)
& 23(2) & -- & 26(3) \\
$\widetilde{\Omega}_{cb}^\star - \widetilde{\Omega}_{cb}$ & 59(8) & 62(13)
& 61(22) & 61(15) \\
$\widetilde{\Xi}_{cb}^\star - \widetilde{\Xi}_{cb}$ & 37(6) & 44(5)
& 44(9) & 42(7)  \\
$\widetilde{\Omega}_b^\star - \widetilde{\Omega}_b$ & 29(5) &
28(11) & 29(4) & 29(7)  \\
$\Omega_b^\prime - \Lambda_b$ & 396(4) & 391(9) & 406(10) & 398(9) \\
$\widetilde{\Xi}_b^\star - \Xi_b$ & 138(20) & 122(38) & 138(46) &
133(36)  \\
$\widetilde{\Xi}_b - \Lambda_b$ & 170(9) & 166(11) & 163(6) & 166(9) \\
$\Lambda_b - B$ & 391(20) & 431(20) & 397(22) & 406(21)  \\
$\widetilde{\Sigma}_b^\star - \widetilde{\Sigma}_b$ & 30(9) &
30(8) & 29(8) & 30(8) \\
$\widetilde{\Sigma}_b^\star - \Lambda_b$ & 224(13) & 203(12) & 175(13)
& 201(13) \\
\end{tabular}
\end{ruledtabular}
\label{tab:one_b_split}
\end{table}

The mass splittings are calculated using ratio of correlators as given
in the Equation (\ref{splitcorr}). As an example, in the Figure
\ref{corrsplitfig} we provide the plots for ratio of correlators,
$\Omega_b^\prime - \Lambda_b$ and $\Xi_b - \Lambda_b$, for comfortable
viewing because of their relatively large mass differences {\em i.e.}
slopes are prominent and well separated. In case of smaller differences,
for instance $\widetilde{\Omega}_{b}^\star - \widetilde{\Omega}_{b}$ or 
$\widetilde{\Xi}_{cb}^\star - \widetilde{\Xi}_{cb}$, the slopes
of the ratio of correlators are rather small and not quite visible. In
the Table \ref{tab:one_b_split} above we collect the results of
single $b$ baryons mass splittings.

\begin{figure}[h]
\scalebox{0.75}{\input{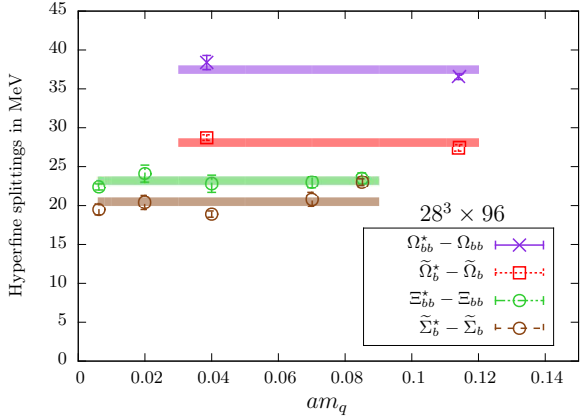}}
\caption{Hyperfine splittings at various $m_s$ and $m_{u/d}$ for a selected
few bottom baryons on $28^3 \times 96$ lattice. Horizontal bands are
the average values of the splittings and is used to guide the eye.}
\label{hsplit_indep_tune}
\end{figure}

Heavy quark basically acts as a static color source, and therefore,
we expect that the hyperfine splittings between states containing single
or multiple  $s$ and $u/d$ quark(s) to depend only weakly on the tuning
of $m_s$ and $m_{u/d}$. For $m_{u/d} \leq 0.085$ and two values of $m_s$
we show this pattern for a few hyperfine splittings in Figure
\ref{hsplit_indep_tune}.
\newline

\noindent
\textbf{Double bottom baryons}: For the baryons containing more than one
$b$-quark, the data are relatively less noisy than those containing
single $b$. The effective mass plots in Figure \ref{efm_b_c}, shown for
only $16^3\times 48$ lattices but similar for two other lattices, is an
evidence for this. The fitting ranges are chosen the same way as is done
for Figure \ref{one_b_efm_demo}.

\begin{figure}[h]
\scalebox{0.75}{\input{efm_l1648.tex}}
\caption{$\Omega_{bbb}^\star$, $\widetilde{\Omega}_{cbb}^\star$
and $\widetilde{\Omega}_{ccb}$ effective masses.}
\label{efm_b_c}
\end{figure}

\noindent
The plot for $\Omega_{bbb}^\star$ appears counter intuitive since being
the heaviest, it is showing lower mass compared to the other two. However,
it receives large correction because of shift in rest mass of three
$b$-quarks.

We tabulate our results for double bottom non-strange baryons in the
Table \ref{tab:multi_b_no_s} while those containing $s$ quark in Table
\ref{tab:multi_b_s}.

\begin{table}[h]
\caption{Triple and double bottom non-strange baryon masses.}
\begin{ruledtabular}
\begin{tabular}{c|c|c|c|c}
Baryon & $16^3\times 48$ & $20^3\times 64$ & $28^3\times 96$ & Average \\
 & (0.15 fm) & (0.12 fm) & (0.09 fm) & (MeV) \\ \hline
$\Omega_{bbb}^\star$ & 1.983(4) & 2.031(3) & 2.154(4) & 14403(7)  \\
$\Omega_{bbb}$ & 1.974(4) & 2.023(5) & 2.148(4) & 14390(8)  \\
$\Omega_{cbb}^\star$ & 2.429(16) & 2.259(4) & 2.117(3) & 11081(21)  \\
$\Omega_{cbb}$ & 2.409(16) & 2.246(5) & 2.110(3) & 11060(23)  \\
$\widetilde{\Omega}_{cbb}^\star$ & 2.431(8) & 2.255(4) & 2.113(3) &
11077(14)  \\
$\widetilde{\Omega}_{cbb}$ &2.432(10) & 2.251(4) & 2.113(3) &
11075(13)  \\
$\Omega_{cbb}^\prime$ & 2.434(8) & 2.250(4) & 2.114(4) &  11076(12) \\
$\Xi_{bb}^\star$ & 1.721(12) & 1.643(10) & 1.666(5) & 10103(24)  \\
$\Xi_{bb}$ & 1.700(12) & 1.640(7) & 1.664(5) & 10091(17)  \\
$\widetilde{\Xi}_{bb}^\star$ & 1.720(12) & 1.635(8) & 1.668(3) &
10100(27) \\
$\widetilde{\Xi}_{bb}$ & 1.703(16) & 1.634(8) & 1.661(4) &
10087(22)  \\
$\Xi_{bb}^\prime$ & 1.704(16) & 1.635(10) & 1.672(3) &
10096(24) \\
\end{tabular}
\end{ruledtabular}
\label{tab:multi_b_no_s}
\end{table}

It is to note that $\Omega_{bbb}$ is a spin-3/2
state having no spin-1/2 counterpart. But in practice we can take a
spin-1/2 projection to get such a fictitious state. Therefore, we label
the physical $(bbb)$ spin-3/2 state with $\Omega_{bbb}^\star$ to keep
consistency with our remaining notation. In this case  none of the
states have PDG entries.

\begin{table}[h]
\caption{Double bottom strange baryon spectra.}
\begin{ruledtabular}
\begin{tabular}{c|c|c|c|c|c}
Baryon & Tuning & $16^3\times 48$ & $20^3\times 64$ & $28^3\times 96$
& Average \\
 && (0.15 fm) & (0.12 fm) & (0.09 fm) & (MeV) \\ \hline
\multirow{2}{*}{$\Omega_{bb}^\star$} & $\eta_s$ & 1.545(11) &
1.536(6) & 1.576(4) & 9902(12) \\
	& $B_s$ & 1.791(12) & 1.703(11) & 1.716(3) & 10203(22) \\ \hline
\multirow{2}{*}{$\Omega_{bb}$} & $\eta_s$ & 1.553(9) &
1.527(7) & 1.570(4) & 9896(13) \\
	& $B_s$ & 1.768(12) & 1.699(8) & 1.715(3) & 10190(17) \\ \hline
\multirow{2}{*}{$\widetilde{\Omega}_{bb}^\star$} & $\eta_s$ & 1.542(9) &
1.529(7) & 1.575(4) & 9896(12) \\
	& $B_s$ & 1.791(12) & 1.693(10) & 1.717(3) & 10199(28) \\ \hline
\multirow{2}{*}{$\widetilde{\Omega}_{bb}$} & $\eta_s$ & 1.541(12) &
1.527(7) & 1.571(4) & 9891(13) \\
	& $B_s$ & 1.789(9) & 1.695(7) & 1.714(3) & 10197(24) \\ \hline
\multirow{2}{*}{$\Omega_{bb}^\prime$} & $\eta_s$ & 1.552(9) &
1.539(7) & 1.578(3) & 9908(11) \\
	& $B_s$ & 1.782(9) & 1.699(7) & 1.720(3) & 10200(20) \\
\end{tabular}
\end{ruledtabular}
\label{tab:multi_b_s}
\end{table}

We would like to point out that the variation of the $\Xi_{bb}\;
(ubb)$ masses with $m_{u/d}$ is almost absent as the major
contribution to these baryons are coming from the two $b$ quarks.
Similarly, from the Table \ref{tab:multi_b_s} we see that the
different tuning of $s$ quark has significantly less influence on
the double bottom baryon masses, a situation unsurprisingly similar
to double bottom baryons with a $u/d$ quark.

The splittings in double bottom sector is tabulated in the Table
\ref{tab:multi_b_split}.
\begin{table}[h]
\caption{Double bottom baryon mass splittings in MeV. None of the
splittings have PDG entries.}
\begin{ruledtabular}
\begin{tabular}{c|c|c|c|c}
Baryon & $16^3 \times 48$ & $20^3\times 64$ & $28^3\times 96$
& Average \\
splittings & (MeV) & (MeV) & (MeV) & (MeV) \\ \hline
$\widetilde{\Omega}_{cbb}^\star - \widetilde{\Omega}_{cbb}$ & -- & 25(5)
& 35(2) & 30(5) \\
$\Omega_{bb}^\star - \Omega_{bb}$ & 34(5) & 25(8) & 37(9) & 32(7) \\
$\Xi_{bb}^\star - \Xi_{bb}$ & -- & 25(4) & 39(7) & 32(5) \\
\end{tabular}
\end{ruledtabular}
\label{tab:multi_b_split}
\end{table}

In the double bottom sector, the splittings between the spin-3/2 and
1/2 states are particularly interesting because HQET relates this mass
differences with hyperfine splittings of heavy-light mesons, which in
the heavy-quark limit \cite{mbbmb}
\begin{equation}
\frac{\Delta M_{bb}^{\text{baryon}}}{\Delta m_b^{\text{meson}}} \;
\rightarrow \; \frac{3}{4}
\end{equation}
This behavior is consistent with our results within errors as can be seen
in Table \ref{tab:mbb_mb_34}.

\begin{table}[h]
\caption{Ratio of hyperfine splittings of doubly heavy baryons to heavy
mesons in the heavy quark limit in $28^3\times 96$ lattice.}
\begin{ruledtabular}
\begin{tabular}{ccccl}
Baryon & Our results & Meson & Our results & Ratio \\
splittings & (MeV) & splittings & (MeV) & \\ \hline
$\tilde{\Omega}_{bbc}^\star - \tilde{\Omega}_{bbc}$ & 35(2) & $B_c^\star
- B_c$ & 46(4) & 0.76(4) \\
$\Omega_{bb}^\star - \Omega_{bb}$ & 37(9) & $B_s^\star - B_s$ & 45(9) &
0.82(9) \\
$\Xi_{bb}^\star - \Xi_{bb}$ & 39(7) & $B^\star - B$ & 47(7) & 0.83(8) \\
\end{tabular}
\end{ruledtabular}
\label{tab:mbb_mb_34}
\end{table}

A few GMO mass relations involving $b$-quark are provided in the reference
\cite{colorint}, which we try to verify in this work,
\begin{eqnarray}
 && M_{\Omega_{ccb}^\star} - M_{\Omega_{ccb}} \approx
M_{\Omega_{cbb}^\star} - M_{\Omega_{cbb}} \label{gmo_ccb_cbb} \\
 && M_{\Sigma_b^\star} - M_{\Sigma_b} \approx
M_{\Xi_{bb}^\star} - M_{\Xi_{bb}} \label{gmo_bb_cc} 
\end{eqnarray}
For the GMO relation (\ref{gmo_ccb_cbb}), the both sides are expected
to be approximately 31 MeV. In our case for $20^3\times 64$ lattice,
for which we have data for both the sides, they are approximately
equal but is around 24 MeV as against 31 MeV given in \cite{colorint}.
Our lattice data is also consistent with the approximate
GMO relation (\ref{gmo_bb_cc}), where each side is about 30 MeV against
20 MeV calculated in \cite{colorint}. 

\section{Summary} \label{sec_sum}
In this paper we presented lattice QCD determination of masses of the
baryons containing one or more $b$ quark(s) using NRQCD action for the
$b$-quark and HISQ action for the $c$, $s$ and $u/d$ quarks. This
combination of NRQCD and HISQ has previously been employed in
\cite{eric} for bottom mesons, however, the exact implementation
was rather different. In this work, we converted the one component
HISQ propagators to $4 \times 4$ matrices by the Kawamoto-Smit
transformation and the two component NRQCD propagators to $4 \times
4$ matrices using the prescription suggested in \cite{brown}.

We have discussed the construction of one and two bottom baryon operators
in details and pointed out the difficulty for constructing
operators motivated by HQET for single bottom baryons. Consequently,
we modified the operators accordingly. For some of the baryons, we have
multiple
operators for the same state {\em i.e.} baryons having the same quantum
numbers. It would be natural in such cases to construct correlation
matrices and obtain lowest lying {\em i.e.} ground states by solving
the generalized eigenvalue method.

\begin{figure}[h]
\scalebox{0.75}{\input{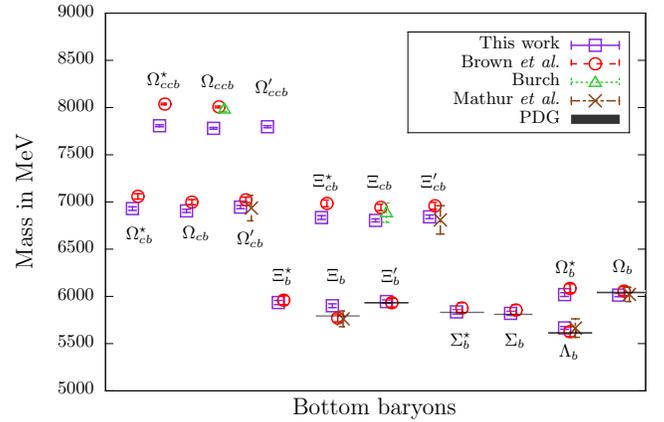}}
\caption{Comparison of our single bottom baryon spectra with Brown {\em
et al.} \cite{brown}, Burch \cite{burch}, Mathur {\em et al.} \cite{mathur}
and PDG \cite{pdg} where available.}
\label{one_b_compare}
\end{figure}

Single bottom baryons can have
isodoublets with the same overall quantum numbers $J^P$. For instance,
there exist three isodoublets of $\Xi_b$ which are not radially or
orbitally excited states \cite{roel}. These states have been
categorized by the spin of the $us$ or $ds$ diquark denoted by $j$
and the spin-parity of the baryon. These baryons are referred to as
$\Xi_b (j=0, J^P={1\over2}^+)$, $\Xi_b^{\prime} (j=1, J^P={1\over2}^+)$
and $\Xi_b^{\star} (j=1, J^P={3\over2}^+)$. The same pattern is
observed in $\Xi_c$ states \cite{pdg}. The mass difference between
$\Xi_b^{\prime}$ and $\Xi_b$ is about 150 Mev. So depending upon
the choice of the wave function having the same overall quantum
numbers we can have different baryon states. If we choose $(s^TC
\gamma_5 d)\,Q$ as our $j=0$ baryon operator then we will be simulating
$\Xi_b$ state and if we project out the spin-1/2 state of $j=1$
operator $(s^TC \gamma_k d)\,Q$ then we will get the $\Xi_b^{\prime}$.
For reason
discussed before, we can not define $j=1$ light-light diquark state.
In our case, the wave function that corresponds to $\Xi_b^{\prime}$
is $(Q^T C\gamma_5 s)\,d$. Constructing operator in this way allows
the $s$ and $d$ quarks to have parallel spin configurations. 
By simple physical reasoning we can argue that explicit construction
of $j=0$ diquark for $\Xi_b$ is more likely to have significant
overlap with physical $\Xi_b$ compared to $\Xi_b^\prime$ ($j=1$) upon
gauge averaging. However the operator $(Q^T C\gamma_5 s)\,d$ is expected 
to have a good overlap with $\Xi_b^{\prime}$ state and this is also
supported by our result. For anti-parallel $s$ and $d$ spin
configuration, $\Xi_b^\prime$ can also have an overlap with the
$\Xi_b$ state. On lattice, operators for states having same quantum
numbers can mix and, therefore, a detailed GEVP analysis can
only resolve the issue of mutual overlap of $\Xi_b$ and $\Xi_b^\prime$
states, which we did not include in this work. This is perhaps the 
reason we see discrepancies in their values with PDG and
others in the Figure \ref{one_b_compare}.

The $b$ mass has been tuned to modified $\Upsilon -\eta_b$ spin
average mass while $c$ quark to $J/\psi -\eta_c$ spin average mass.
The $s$ quark required to be tuned to both the fictitious $\eta_s$
and $B_s$ mass since we expect the bottom-strange bound state to be
more appropriate than $s-\bar{s}$ bound state in bottom baryons.
For the light $u/d$ quarks, we have considered a wide range of bare
masses and tune it using $B$ meson. 
However, this scheme of tuning
$u/d$ quarks has not worked for $\Lambda_b$. There $u/d$ are tuned
to capture the 150 MeV mass difference $\Sigma_b - \Lambda_b$. 
This specially tuned $m^\prime_{u/d}$, which is used only for 
$\Lambda_b$, gives it a mass of 5667 MeV. The PDG value for $\Lambda_b$ mass
is 5620 MeV. We demonstrated the
variation of bottom baryons as well as hyperfine splittings against
varying $m_s$ and $m_{u/d}$. We showed that the hyperfine splittings
are almost independent of $s$ and $u/d$ quark masses.

\begin{figure}[h]
\scalebox{0.75}{\input{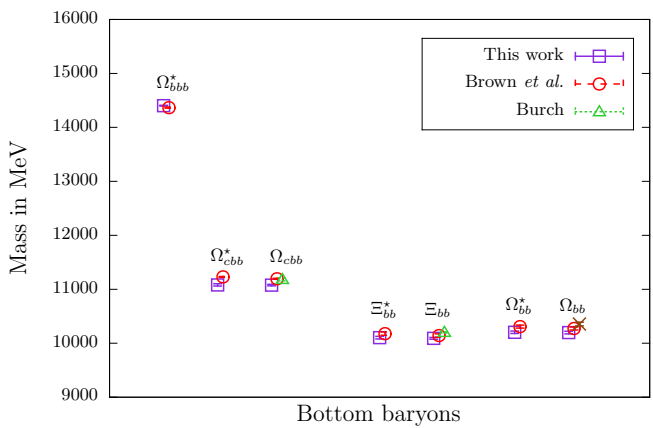}}
\caption{Comparison of our triple and double bottom baryon spectra with
Brown {\em et al.} \cite{brown} and Burch \cite{burch}.}
\label{two_b_compare}
\end{figure}

We compare our bottom baryon results with other works, mostly with
\cite{mathur, burch, brown}, in the Figures \ref{one_b_compare} and
\ref{two_b_compare}. NRQCD has been standard action of choice for
$b$ quark in these three cited studies, but the actions used for
$c$ are all different -- NRQCD, Clover-Wilson and relativistic heavy
quark action \cite{tsukuba}. Whatever differences we see in the
results for single $b$ baryons with $c$ quark, particularly in
the cases with two $c$, possibly have stemmed from the differences
in actions. However, in this work we do not address the systematics 
involved, which could be significant, because of these differences. 
(The study of such systematics needed to arrive at phenomenologically 
relevant numbers will be reported elsewhere.)
But otherwise, the results of bottom baryon spectra in
the present study with NRQCD $b$ quark and HISQ $c,s,u/d$ quarks
appear to agree with each other. We would like to emphasis again 
that the errors we reported here are only statistical.

The comparison of the hyperfine splittings is shown in the Figure
\ref{bsplit_compare_h}.

\begin{figure}[h]
\scalebox{0.75}{\input{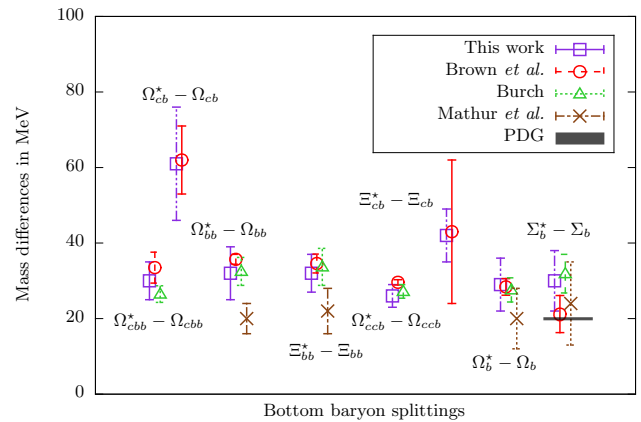}}
\caption{Hyperfine splittings of bottom baryons calculated in this
work and compared with Brown {\em et al.} \cite{brown}, Burch \cite{burch},
Mathur {\em et al.} \cite{mathur} and PDG \cite{pdg} where available.}
\label{bsplit_compare_h}
\end{figure}

Apart from the hyperfine splittings, a few other mass splittings
calculated in this work are assembled in the Table \ref{tab:mass_diff}.
The bottom baryon spectra and various mass splittings reported in
this paper and those appearing in \cite{brown, pdg} are
well comparable given the wide choice of actions and tuning employed
in achieving them. 

\begin{table}[h]
\caption{Bottom baryon mass differences in MeV. PDG values without error
is simply the differences of the two states.}
\begin{ruledtabular}
\begin{tabular}{cccc}
Mass & This work & Brown {\em et al.} & PDG \\
splittings & & \cite{brown} & \cite{pdg} \\ \hline
$\Omega_b^\prime - \Lambda_b$ & 398(9) & -- & 426.4(2.2) \\
$\Xi_b^\star - \Xi_b$ & 133(36) & 189(29) & 155.5 \\
$\Xi_b - \Lambda_b$ & 166(9) & -- & 172.5(0.4) \\
$\Lambda_b - B$ & 406(21) & -- & 339.2(1.4) \\
$\Sigma_b^\star - \Lambda_b$ & 201(13) & 251(46) & 213.5 \\
\end{tabular}
\end{ruledtabular}
\label{tab:mass_diff}
\end{table}

\section{Acknowledgement}
The numerical part of this work has been performed at HPC facility in
NISER (Kalinga cluster) funded by Dept. of Atomic Energy (DAE), Govt.
of India. A significant part of this work has been carried out in
Proton cluster funded by DST-SERB project number SR/S2/HEP-0025/2010.
The authors acknowledge useful discussions with Dipankar Chakrabarti
(IIT-Kanpur, India) and Stefan Meinel (University of Arizona, USA)
on bottom baryon operator construction. One of the authors (PM) thanks
DAE for financial support.

\newpage
\onecolumngrid
\appendix*\section{Bottom baryon masses in MeV}

\noindent
For the reason of completeness, we provide the bottom baryon masses
in MeV for different lattices used in this work. In these tables 
we provide only $B_s$ tuned strange bottom baryon masses and not
$\eta_s$ tuned values.

\begin{table}[h]
\caption{Single bottom baryon masses in MeV, corresponding
to the Tables \ref{tab:one_b_no_s} and \ref{tab:one_b_s}.}
\begin{ruledtabular}
\begin{tabular}{ccccc||ccccc}
Baryons & $16^3\times 48$ & $20^3\times 64$ & $28^3\times 96$
& Average & Baryons & $16^3\times 48$ & $20^3\times 64$ & $28^3\times
96$ & Average \\
 & (0.15 fm) & (0.12 fm) & (0.09 fm) & & & (0.15 fm) & (0.12 fm) &
(0.09 fm) & \\ \hline
$\widetilde{\Omega}_{ccb}^\star$ & 7816(6) & 7794(6) & 7805(6) & 7807(11) &
{$\widetilde{\Omega}_{cb}^\star$} & 6945(9) & 6906(10) & 6938(9) & 6930(19) \\
$\widetilde{\Omega}_{ccb}$ & 7788(6) & 7769(5) & 7782(6) & 7780(9) &
{$\Omega_{cb}$} & 6887(14) & 6873(11) & 6919(4) & 6893(16) \\
$\Omega_{ccb}^\prime$ & 7813(5) & 7794(5) & 7782(6) & 7797(11) &
{$\widetilde{\Omega}_{cb}$} & 6912(10) & 6883(11) & 6923(4) & 6906(17) \\
$\widetilde{\Xi}_{cb}^\star$ & 6853(8) & 6810(6) & 6840(13) & 6835(20) &
{$\Omega_{cb}^\prime$} & 6951(8) & 6921(10) & 6965(4) & 6946(17) \\
$\Xi_{cb}$ & 6794(14) & 6781(6) & 6785(11) & 6787(12) &
{$\widetilde{\Xi}_b^\star$} & 5957(10) & 5911(10) & 5934(13) & 5935(22) \\
$\widetilde{\Xi}_{cb}$ & 6822(10) & 6789(6) & 6802(9) & 6805(16) &
{$\Xi_b$} & 5929(12) & 5900(5) & 5873(13) & 5901(20) \\
$\Xi_{cb}^\prime$ & 6859(8) & 6819(6) & 6851(13) & 6843(19) &
{$\widetilde{\Xi}_b$} & 5945(14) & 5908(7) & 5910(9) & 5921(19) \\
$\widetilde{\Sigma}_b^\star$ & 5861(10) & 5812(8) & 5833(11) & 5836(22) &
{$\Xi_b^\prime$} & 5961(13) & 5934(7) & 5941(13) & 5946(16) \\
$\widetilde{\Sigma}_b$ & 5850(9) & 5809(5) & 5800(13) & 5820(21) &
{$\widetilde{\Omega}_{b}^\star$} & 6049(10) & 6010(10) & 5998(7) & 6019(20) \\
$\Sigma_b^\prime$ & 5864(9) & 5834(5) & 5844(20) & 5848(18) &
{$\widetilde{\Omega}_{b}$} & 6035(14) & 6008(11) & 5998(7) & 6014(17) \\
$\Lambda_b$ & 5669(9) & 5674(10) & 5658(20) & 5667(14) &
{$\Omega_{b}^\prime$} & 6054(10) & 6031(11) & 6066(4) & 6051(15)
\end{tabular}
\end{ruledtabular}
\label{tab:one_b_mev}
\end{table}

\begin{table}[h]
\caption{Triple and double bottom baryon masses in MeV,
corresponding to the Tables \ref{tab:multi_b_no_s} and \ref{tab:multi_b_s}.}
\begin{ruledtabular}
\begin{tabular}{ccccc||ccccc}
Baryons & $16^3\times 48$ & $20^3\times 64$ & $28^3\times 96$
& Average & Baryons & $16^3\times 48$ & $20^3\times 64$ & $28^3\times
96$ & Average \\
 & (0.15 fm) & (0.12 fm) & (0.09 fm) & & & (0.15 fm) & (0.12 fm) &
(0.09 fm) & \\ \hline
$\Omega_{bbb}^\star$ & 14399(5) & 14405(5) & 14403(9) & 14403(7) &
{$\Omega_{bb}^\star$} & 10217(16) & 10177(18) & 10216(7) & 10203(22) \\
$\Omega_{bbb}$ & 14388(5) & 14392(8) & 14390(9) & 14390(8) &
{$\Omega_{bb}$} & 10187(16) & 10171(13) & 10214(7) & 10190(17) \\
$\Omega_{cbb}^\star$ & 11056(21) & 11091(6) & 11095(7) & 11081(21) &
{$\widetilde{\Omega}_{bb}^\star$} & 10217(16) & 10161(16) & 10218(7) &
10199(28) \\
$\Omega_{cbb}$ & 11029(21) & 11070(8) & 11080(7) & 11060(23) &
{$\widetilde{\Omega}_{bb}$} & 10214(12) & 10164(11) & 10212(7) & 10197(24) \\
$\widetilde{\Omega}_{cbb}^\star$ & 11058(10) & 11085(7) & 11086(7) &
11077(14) &
{$\Omega_{bb}^\prime$} & 10205(12) & 10171(11) & 10225(7) & 10200(20) \\
$\widetilde{\Omega}_{cbb}$ & 11060(13) & 11078(7) & 11086(7) & 11075(13)
 & & & & & \\
$\Omega_{cbb}^\prime$ & 11062(10) & 11076(7) & 11088(9) & 11076(12)
 & & & & & \\
$\Xi_{bb}^\star$ & 10124(16) & 10078(16) & 10106(11) & 10103(24)
 & & & & & \\
$\Xi_{bb}$ & 10097(16) & 10073(11) & 10102(11) & 10091(17)
 & & & & & \\
$\widetilde{\Xi}_{bb}^\star$ & 10123(16) & 10065(13) & 10111(7) & 10100(27)
 & & & & & \\
$\widetilde{\Xi}_{bb}$ & 10101(21) & 10063(13) & 10095(9) & 10087(22)
 & & & & & \\
$\Xi_{bb}^\prime$ & 10102(21) & 10065(16) & 10119(7) & 10096(24)
 & & & & & \\
\end{tabular}
\end{ruledtabular}
\label{tab:multi_b_no_s_mev}
\end{table}

\twocolumngrid


\end{document}